\documentclass[aps,twocolumn,showpacs,pra,amsmath,amssymb,floatfix,superscriptaddress]{revtex4-2}
\usepackage{amsmath,amssymb,graphicx,color}
\usepackage{bm}
\usepackage{bbm}
\usepackage[caption=false]{subfig}  
\usepackage[usenames,dvipsnames]{xcolor}
\usepackage[normalem]{ulem}
\usepackage{soul}
\usepackage[colorlinks=true,citecolor=Cerulean,linkcolor=RubineRed,urlcolor=Cerulean]{hyperref}
\usepackage{soul,xcolor}

\newcommand{\Tr}{\ensuremath{\mathrm{Tr}\,}}

\usepackage{mathptmx} 

\usepackage{braket}

\begin{document}
\title{Aplication of simultaneous and continuous measurement of noncommutative observables: Preparation of the pure ideal quadrature-squeezed state by feedback control}

\author{Chao Jiang} \email{chaojiang@zju.edu.cn}
\affiliation{Department of Physics and Zhejiang Institute of Modern Physics, Zhejiang University, Hangzhou, Zhejiang 310027, China}

\author{Gentaro Watanabe}\email{gentaro@zju.edu.cn}
\affiliation{Department of Physics and Zhejiang Institute of Modern Physics, Zhejiang University, Hangzhou, Zhejiang 310027, China}
\affiliation{Zhejiang Province Key Laboratory of Quantum Technology and Device, Zhejiang University, Hangzhou, Zhejiang 310027, China}

\begin{abstract}
As an application of the simultaneous and continuous measurement of noncommutative observables formulated in our previous paper [C. Jiang and G. Watanabe, \href{https://doi.org/10.1103/PhysRevA.102.062216}{Phys.\ Rev.\ A {\bf 102}, 062216 (2020)}], we propose a scheme to generate a pure ideal quadrature-squeezed state in a one-dimensional harmonic oscillator system by the feedback control based on such type of measurement of noncommutative quadrature observables. We find that, by appropriately setting the strengths of the measurement and the feedback control, the pure ideal quadrature-squeezed state with arbitrary squeezedness can be produced. This is in contrast to the scheme based on the single-observable measurement and the feedback control, where only nonideal squeezed states are produced. Furthermore, we discuss the transient dynamics of the system and the experimental feasibility of our scheme.

\end{abstract}

\maketitle

\section{Introduction \label{sec:intro}}  

The squeezed state \cite{M. O. Scully, C. C. Gerry, D. F. Walls, R. Loudon, V. V. Dodonov, U. L. Andersen, R. Schnabel} is a kind of nonclassical state exhibiting many unique  properties, which cannot be seen in the coherent state \cite{M. O. Scully, C. C. Gerry, R. Loudon, V. V. Dodonov, R. Schnabel, D. F. Walls2, H. P. Yuen, L. Mandel, R. Short, G.J. Milburn, J. P}. In previous decades, there has been a dramatic development of the theoretical \cite{R. Simon, D. J. Wineland, R. G., L. P., T. Kiesel} and experimental \cite{J. Jing, R. Dong, M. S. Stefszky, T. Eberle, H. Vahlbruch}  studies on the squeezed state. These explorations enable people not only to understand the nature of quantum mechanics more deeply, but also to find extensive applications of squeezed states in various situations, such as reducing the noise in the quantum communication \cite{H. P. Yuen2, R. E. Slusher1}, improving the sensitivity of the interferometers to realize more precise measurement \cite{C. M. Caves, LIGO Scientific Collaboration, J. Aasiand LIGO collaborators}, and enhancing the performance of the quantum heat engines \cite{J. Rossnagel}.

The first experimental realization of the squeezed light was achieved by Slusher and his collaborators by four-wave mixing in an optical cavity \cite{R. E. Slusher}. In addition, several other practical and effective schemes to generate the squeezed states have also been proposed and developed in previous decades \cite{D. F. Walls, G. Milburn2, W. Becker, L. Mandel2, B. Yurke, Yamamoto}. Among them, the feedback control is a powerful and commonly used technique \cite{Yamamoto, H. M. Wiseman, A. C. Doherty, L. K. Thomsen, K. C. Cox}. Based on the measurement signals obtained from the system, the feedback control scheme allows us to manipulate the evolution of the system in a robust manner and drive it to the target squeezed state.

Nevertheless, most of the feedback control schemes used to generate the squeezed state are based on the single-observable measurement \cite{H. M. Wiseman, A. C. Doherty, L. K. Thomsen,  K. C. Cox}, and only a few works have considered the two-observable measurement case \cite{A. Levy}. Actually, the multiobservable measurement can gain more information about the system, and the feedback control based on it offers a more flexible way to prepare the desired squeezed state. In the present paper, we will specifically consider the case where two observables of the system are simultaneously measured. The issues of the simultaneous and continuous measurement of two noncommutative observables have already been addressed in our previous work \cite{C. Jiang}, and the evolution equations of the system under the measurement, i.e., the It\^{o} stochastic equation and the Lindblad form master equation, have also been derived there \cite{C. Jiang}. In the present paper, as an important application of our previous works, we shall propose a scheme to generate the quadrature-squeezed state by the feedback control approach based on the simultaneous and continuous measurement.  We will show that by properly setting the feedback control Hamiltonian, any minimum uncertainty quadrature-squeezed state of the system can be obtained after sufficiently long, but experimentally feasible time evolution. These squeezed states have important applications in, e.g., metrology and quantum thermodynamics.

As a simple but important example of the quadrature squeezing, we take a harmonic oscillator model, which provides a good description of various systems such as a photon field in a single-mode optical cavity as well as the microvibration of a trapped atom. Recently, experiments using levitated particles in vacuum have made many outstanding achievements in exploring the motion of the quantum systems \cite{J. Millen, U. Deli, M. Rashid, T. M. Hoang, M. Rashid, S. Kuhn, G. P. Conangla} due to the rapid development of the optomechanics and levitodynamics \cite{M. Aspelmeyer, J. Millen2, C. G.-Ballestero}. By loading the particles into an optical or magnetic trap in vacuum, the system is extremely decoupled from the environment and the oscillation of the particles can be controlled with high precision via tuning the trapping potential \cite{J. Millen, U. Deli, T. M. Hoang, M. Rashid, S. Kuhn, G. P. Conangla}. To keep the generality of the discussion, we do not specify a particular physical setup of the system in the beginning of the paper. We consider the simultaneous and continuous measurement of two quadratures of the harmonic oscillator system and model the evolution of the system by the master equations derived in our previous paper \cite{C. Jiang}. Then, we work out the time evolution of the variances of the quadratures, and show that any minimum uncertainty quadrature-squeezed state can be obtained as the asymptotic steady state by setting the proper feedback control strength. It is noted  that we can create the ideal quadrature-squeezed state with arbitrary high squeezing in principle, which cannot be realized for the feedback control scheme based on the single-observable measurement. Finally, we verify our results by numerical simulations and show that the ideal quadrature-squeezed state can be created by our scheme within a typical time scale of the experiments of levitated particles.

The structure of our paper is as follows. In Sec.~\ref{sec:model}, we derive the master equation of the harmonic oscillator system under the simultaneous, continuous measurement and the feedback control. In Sec.~\ref{sec:disc}, we work out the time evolution equations of the uncertainties of two quadratures, and find the condition under which the arbitrary pure ideal quadrature-squeezed state can be generated. Moreover, the comparison between our scheme and the single-observable measurement case is also given in this section. The numerical demonstrations and the experimental feasibility are presented in Sec.~\ref{sec:results}. Section \ref{sec:conclusion} concludes our paper and gives future prospects of the scheme.

\section{measurement and feedback control model \label{sec:model}}

We consider a one-dimensional harmonic oscillator. For photons in a single-mode optical cavity with frequency $\omega$, the Hamiltonian of the system $\hat{H}_s$ is given by
\begin{align}
\hat{H}_s  = \hbar \omega \left( \hat{c}^{\dagger} \hat{c} + \frac{1}{2} \right),
\label{eq:photon}
\end{align}
where $\hat{c}^{\dagger}$ and $\hat{c}$ are the creation and annihilation operators of a photon, respectively. For simplicity, we will set $\hbar = \omega = 1$ throughout the remaining part of the paper. By introducing the following two quadrature operators of the system, 
\begin{align}
	\hat{x} = \frac{1}{\sqrt{2}} \left( \hat{c}^{\dagger} + \hat{c} \right) ,
	\\
	\hat{p} = \frac{i}{\sqrt{2}} \left( \hat{c}^{\dagger} - \hat{c} \right) ,
\end{align}
the Hamiltonian can be rewritten as
\begin{align}
\hat{H}_s  =  \frac{1}{2} \left(  \hat{p}^2  + \hat{x}^2 \right).  
\label{eq:oscillator}
\end{align}
For mechanical harmonic oscillator systems, $x$ and $p$ quadratures correspond to the position and the momentum of the oscillator, respectively, and thus the Hamiltonian can be divided into the kinetic energy part and the potential energy part. For the convenience of the discussion, we use both representations of Eqs.~(\ref{eq:photon}) and (\ref{eq:oscillator}) in the following analysis.

We consider the situation in which $\hat{x}$ and $\hat{p}$ are simultaneously and continuously measured. The state of the system at time $t$ is denoted by $\hat{\rho}(t)$, and the conditioned master equation of the system under the measurement is given by \cite{A. Levy, C. Jiang, A. J. Scott}
\begin{align}
	d\hat{\rho}= & -i[\hat{H}_s, \hat{\rho}]\, dt -\frac{\gamma_x}{8}\boldsymbol{[}\hat{x}, [\hat{x}, \hat{\rho}]\boldsymbol{]}\, dt - \frac{\gamma_p}{8}\boldsymbol{[}\hat{p}, [\hat{p}, \hat{\rho}]\boldsymbol{]}\, dt 
	\notag
	\\
	& +\sqrt{\gamma_x}\, \mathcal{H} \left[  \left(  \hat{x} - \left \langle \hat{x} \right \rangle  \right) \hat{\rho} \right] \, d\xi_x 
	\notag
	\\
	& + \sqrt{\gamma_p}\, \mathcal{H} \left[ \left( \hat{p} - \left\langle  \hat{p} \right\rangle  \right) \hat{\rho} \right] \, d\xi_p.
	\label{eq:cme} 
\end{align}
Here, $\gamma_j > 0$ is the strength of the measurement of $\hat{x}$ $(\mbox{for } j = x)$ and $\hat{p}$ $(\mbox{for } j = p)$, $\left \langle \hat{A} \right \rangle$ is the average of the observable $\hat{A}$, which is defined as
\begin{align}
\left \langle \hat{A} \right \rangle \equiv \Tr{\left( \hat{A} \hat{\rho} \right) },
\end{align}
the symbol $\mathcal{H}[\hat{O}]$ is the Hermitian part of operator $\hat{O}$:
\begin{align}
\mathcal{H}[\hat{O}]  \equiv \frac{1}{2}(\hat{O} + \hat{O}^\dagger),
\end{align}
and $d\xi_j$'s $(j = x \mbox{\ and \ } p)$ are independent It\^{o} increments satisfying \cite{C. W., N. G. V. Kampen, note:noise} 
\begin{align}
& E[d\xi_j] = 0, 
\\
& E[d\xi_x(t) \cdot d\xi_p(t)] = 0,
\\
& E[d\xi_j(t) \cdot d\xi_j(t)] = dt.
\end{align}

This It\^{o} stochastic master equation (\ref{eq:cme}) can be derived from a particular measurement model by Arthurs and Kelly \cite{C. Jiang, A. J. Scott, E. Arthurs} within the Born-Markov approximation \cite{H.-P. Breuer, H. M. W., K. Jacobs2}. It is noted that even though the simultaneously measured quantities $\hat{x}$ and $\hat{p}$ are noncommutative, the master equation (\ref{eq:cme}) obtained in the Born-Markov approximation does not contain a cross term of these two measurements \cite{C. Jiang}. Moreover, although the underlying measurement model by Arthurs and Kelly is schematic, the resulting master equation (\ref{eq:cme}) is widely used in theoretical studies \cite{A. Levy, C. Jiang, A. J. Scott, H. M. W., K. Jacobs2, Luis, A. Chantasri} and the analysis of experimental results \cite{Shay, A. Setter, M. Rossi, L. S. Walker}.

By taking the ensemble average of Eq.~(\ref{eq:cme}), we readily obtain the following unconditioned master equation:
\begin{align}
\frac{d\hat{\rho}}{dt}= & -i[\hat{H}_s, \hat{\rho}] -\frac{\gamma_x}{8}\boldsymbol{[}\hat{x}, [\hat{x}, \hat{\rho}]\boldsymbol{]} - \frac{\gamma_p}{8}\boldsymbol{[}\hat{p}, [\hat{p}, \hat{\rho}]\boldsymbol{]}
\notag
\\
= & -i \left[ \hat{H}_s, \hat{\rho} \right] + \frac{\gamma_p}{4} \left( \mathcal{D}[\hat{c}] +  \mathcal{D}[\hat{c}^{\dagger}] \right) \hat{\rho} 
\notag
\\
 & + \frac{\gamma_x - \gamma_p}{8} \mathcal{D}[\hat{c} + \hat{c}^{\dagger}] \hat{\rho},
\label{eq:ucme} 
\end{align}
where the superoperator $\mathcal{D}[\hat{O}]$ is defined for an arbitrary operator $\hat{O}$ as 
\begin{align}
\mathcal{D}[\hat{O}] \, \hat{\rho} \equiv \hat{O} \hat{\rho}  \hat{O}^{\dagger} - \frac{1}{2}\left(\hat{O}^{\dagger} \hat{O} \hat{\rho}   +  \hat{\rho}  \hat{O}^{\dagger} \hat{O}\right).
\end{align}	

The first term in the right-hand side of Eq.~(\ref{eq:ucme}) represents the unitary evolution governed by the Hamiltonian, while the remaining terms represent the effects of the measurement. To get more insights into the measurement effects, we consider the change of the average of the kinetic and potential energies induced by the measurement, which is given by 
\begin{align}
\frac{d \left\langle \hat{p}^2 \right\rangle }{dt} = \frac{\gamma_x}{4} > 0,
\label{eq:increT}
\\
\frac{d \left\langle \hat{x}^2 \right\rangle }{dt} = \frac{\gamma_p}{4} > 0.
\label{eq:increV}
\end{align}
Notice that the measurement of $\hat{x}$ leads to the increment of the average of the kinetic energy, and its increase rate is larger for larger measurement strength $\gamma_x$, while the measurement of $\hat{p}$ yields the similar results for the average of the potential energy. 

Equations (\ref{eq:increT}) and (\ref{eq:increV}) show that the system keeps on getting the energy by the continuous measurement, and thus it never reaches the steady state. This result can easily be understood from Eq.~(\ref{eq:ucme}): The term $\left( \mathcal{D}[\hat{c}] +  \mathcal{D}[\hat{c}^{\dagger}] \right) \hat{\rho}$ in the equation represents that the system is effectively connected to a heat bath with infinite temperature, which leads to the divergence of the asymptotic internal energy of the system. 

In order to regulate the energy of the system to be finite, we perform the feedback control based on the outcome of the continuous measurement. As will be clarified later in Sec.~\ref{sec:disc}, to obtain squeezed states, we take the feedback control Hamiltonian $\hat{H}_f$ in the following form \cite{A. Levy}:
\begin{align}
\hat{H}_f dt \equiv  -\kappa_f \bar{x}(t)dt \cdot \hat{p} + \kappa_f \bar{p}(t)dt \cdot \hat{x},  \label{eq:fb}
\end{align}
where $\kappa_f$ is a real positive parameter called feedback control strength, and $\bar{x}(t)dt$ and $\bar{p}(t)dt$ are measurement signals defined as \cite{A. Levy}
\begin{align}
\bar{x}(t)dt \equiv \left\langle \hat{x} \right\rangle dt + \frac{d \xi_x}{\sqrt{\gamma_x}},
\\
\bar{p}(t)dt \equiv \left\langle \hat{p} \right\rangle dt + \frac{d \xi_p}{\sqrt{\gamma_p}}.
\end{align}
We now further assume that the delay time of the feedback control is short enough such that the total process can be approximated by a Markovian process \cite{H. M. Wiseman}, and then the state of the system under the simultaneous, continuous measurement and the feedback control becomes $\exp(-i\hat{H}_f dt )\, (\hat{\rho} + d\hat{\rho})\, \exp(i\hat{H}_f dt)$. By applying the Baker-Campbell-Hausdorff formula and the It\^{o} rule, and keeping all the terms up to the first order of $dt$, we have only four terms: $(\hat{\rho} + d\hat{\rho})$, $-i [\hat{H}_f dt, \hat{\rho}]$, $-i [\hat{H}_f dt, d \hat{\rho}]$, and $-2^{-1} [\hat{H}_f dt, [\hat{H}_f dt, \hat{\rho} ]]$. The resulting master equation of the system after taking the ensemble average reads \cite{C. Jiang, note:term} 

\begin{align}
\frac{d \hat{\rho}}{dt} = -i \left[ \hat{H}_s, \hat{\rho} \right] + k_1 \mathcal{D}[\hat{c}] \hat{\rho} + k_2 \mathcal{D}[\hat{c}^{\dagger}] \hat{\rho}  + k_3 \mathcal{D}[\hat{c} + \hat{c}^{\dagger}] \hat{\rho},    \label{eq:me}
\end{align}
with
\begin{align}
k_1 & \equiv \frac{\gamma_p}{4} + \frac{\kappa_f^2}{\gamma_x} + \kappa_f, \label{eq:k1}
\\
k_2 & \equiv \frac{\gamma_p}{4} + \frac{\kappa_f^2}{\gamma_x} - \kappa_f, \label{eq:k2}
\\
k_3 & \equiv \frac{\gamma_x - \gamma_p}{8} - \frac{\kappa_f^2}{2\gamma_x} + \frac{\kappa_f^2}{2\gamma_p}.  \label{eq:k3}
\end{align}

Equations (\ref{eq:ucme}) and (\ref{eq:me}) share the similar structure, and the former master equation for the case without feedback control can easily be obtained by simply setting $\kappa_f = 0$ in Eq.~(\ref{eq:me}). Comparing these two equations, we find that there are two kinds of additional terms in Eq.~(\ref{eq:me}) resulting from the feedback control. One is the term whose coefficients are proportional to $\kappa_f^2$, and the other is the term whose coefficients are proportional to $\kappa_f$. Let us first focus on the effect of the former kind of additional term, which comes from the fluctuations of the measurement signals, i.e., the term $-2^{-1} [\hat{H}_f dt, [\hat{H}_f dt, \hat{\rho} ]]$, and temporarily ignore the latter term. After introducing the following two parameters $\Gamma_x$ and $\Gamma_p$,
\begin{align}
\Gamma_x & \equiv \gamma_x  + \frac{4\kappa_f^2}{\gamma_p} ,   \label{eq:Gammax}
\\
\Gamma_p & \equiv \gamma_p  + \frac{4\kappa_f^2}{\gamma_x} ,   \label{eq:Gammap}
\end{align}
the master equation (\ref{eq:me}) without the additional terms proportional to $\kappa_f$ can be rewritten as 
\begin{align}
\frac{d\hat{\rho}}{dt} = & -i \left[ \hat{H}_s, \hat{\rho} \right] + \frac{\Gamma_p}{4} \left( \mathcal{D}[\hat{c}] +  \mathcal{D}[\hat{c}^{\dagger}] \right) \hat{\rho} 
\notag
\\
& + \frac{\Gamma_x - \Gamma_p}{8} \mathcal{D}[\hat{c} + \hat{c}^{\dagger}] \hat{\rho}.
\label{eq:effucme} 
\end{align}
By comparing Eqs.~(\ref{eq:ucme}) and (\ref{eq:effucme}), we can see that $\Gamma_x$ and $\Gamma_p$ play a role of effective measurement strengths of $\hat{x}$ and $\hat{p}$, respectively. Hence, as can be seen from Eqs.~(\ref{eq:Gammax}) and (\ref{eq:Gammap}), the effect of $\kappa_f^2$ terms is to enhance the increase rate of the internal energy of the system according to our previous discussion.

Next, we focus on the terms proportional to $\kappa_f$: $\kappa_f \big( \mathcal{D}[\hat{c}] - \mathcal{D}[\hat{c}^{\dagger}] \big) \hat{\rho}$. Note that they originate from the term $-i [\hat{H}_f dt, d \hat{\rho}]$, which represents the effect of the interplay between the noise of the measurement outcome and the feedback control signals. These additional terms reduce the energy of the system since
\begin{align}
\Tr{\left\lbrace \hat{p}^2  \kappa_f \left( \mathcal{D}[\hat{c}] - \mathcal{D}[\hat{c}^{\dagger}] \right) \hat{\rho} \right\rbrace } = -2\kappa_f \left\langle \hat{p}^2 \right\rangle < 0,   \label{eq:extractT}
\\
\Tr{\left\lbrace \hat{x}^2   \kappa_f \left( \mathcal{D}[\hat{c}] - \mathcal{D}[\hat{c}^{\dagger}] \right) \hat{\rho} \right\rbrace } = -2\kappa_f \left\langle \hat{x}^2 \right\rangle < 0.   \label{eq:extractV}
\end{align}
Because of these terms, the energy increase by the continuous measurement can be balanced with the energy reduction by the feedback control, so that the system can reach a steady state after a sufficiently long time.

\section{Discussion \label{sec:disc}}

Before proceeding, let us briefly review the definition of squeezed states and ideal squeezed states \cite{M. O. Scully, C. C. Gerry, D. F. Walls}. For two arbitrary Hermitian operators $\hat{A}$ and $\hat{B}$, the product of the uncertainties of the operators, $\left\langle \Delta \hat{A}^2 \right\rangle \left\langle \Delta \hat{B}^2 \right\rangle $ with $\Delta \hat{A} \equiv \hat{A} - \left\langle \hat{A} \right\rangle $ and $\Delta \hat{B} \equiv \hat{B} - \left\langle \hat{B} \right\rangle $, satisfies 
\begin{align}
\left\langle \Delta \hat{A}^2 \right\rangle \left\langle \Delta \hat{B}^2 \right\rangle \geqslant \frac{1}{4} \left| \left\langle [\hat{A}, \hat{B}]\right\rangle \right| ^2 
\label{eq:uncer}
\end{align}
according to the Heisenberg uncertainty principle. The state of the system is a squeezed state if either 
\begin{align}
	\left\langle \Delta \hat{A}^2 \right\rangle < \frac{1}{2} \left| \left\langle [\hat{A}, \hat{B}] \right\rangle \right| 
\end{align}
or
\begin{align}
\left\langle \Delta \hat{B}^2 \right\rangle < \frac{1}{2} \left| \left\langle [\hat{A}, \hat{B}] \right\rangle \right| 
\end{align}
is satisfied. In particular, squeezed states for which the equality in  Eq.~(\ref{eq:uncer}) holds are ideal squeezed states.

Squeezing implies various physical meanings depending on the system considered.  Taking the harmonic oscillator as an example, the quadrature squeezing for photons in a single-mode cavity means that the noise in the corresponding quadrature is reduced below that of the coherent state. On the other hand, for an atom trapped in a harmonic oscillator potential, the squeezing of the $x$ quadrature means the localization or confinement of the atom in the position space, while the squeezing of the $p$ quadrature implies that the atom is cooled \cite{A. C. Doherty}. 

Here, we propose a scheme to prepare quadrature-squeezed states based on the simultaneous, continuous measurement and the feedback control introduced in the previous section. In our proposal, the target squeezed state is obtained as an asymptotic steady state of the system. To study the condition under which the squeezed states can be generated,  we mainly focus on the first and the second moments of observables $\hat{x}$ and $\hat{p}$, since the uncertainties are determined only by these two moments. In order to get clear understanding of the effects of the measurement and feedback control by themselves, we first ignore the unitary evolution term $-i [\hat{H}_s, \hat{\rho}]$. This is valid when the effects of the measurement and the feedback control is predominant compared to the unitary evolution: For instance, the absolute values of the coefficients $k_1$, $k_2$, and $k_3$ are much larger than unity \cite{note:setting}. The effect of the unitary evolution term will be briefly discussed later in Sec.~\ref{sec:results}.

We first consider the time evolution of the average of $\hat{x}$ and $\hat{p}$, which can easily be obtained from Eq.~(\ref{eq:me}):
\begin{align}
\frac{d \left\langle \hat{x} \right\rangle }{dt} & = \frac{k_2 - k_1}{2} \left\langle \hat{x} \right\rangle = -\kappa_f \left\langle \hat{x} \right\rangle,  \label{eq:xdot}
\\
\frac{d \left\langle \hat{p} \right\rangle }{dt} & = \frac{k_2 - k_1}{2} \left\langle \hat{p} \right\rangle = -\kappa_f \left\langle \hat{p} \right\rangle.  \label{eq:pdot}
\end{align}
These two averages reach the steady values exponentially in time, and the steady solutions of Eqs.~(\ref{eq:xdot}) and (\ref{eq:pdot}) are
\begin{align}
\left\langle \hat{x}\right\rangle = 0,
\\
\left\langle \hat{p}\right\rangle = 0,
\end{align} 
respectively. Therefore, the uncertainties of $\hat{x}$ and $\hat{p}$, $\left\langle \Delta \hat{x}^2 \right\rangle $ and $\left\langle \Delta \hat{p}^2 \right\rangle $, are determined solely by the second moments of $\hat{x}$ and $\hat{p}$, or in other words, the averages of the potential and the kinetic energies of the system, respectively. This provides us another point of view on the preparation of quadrature squeezed states as the control of the energy. As shown in the previous section, the kinetic energy and the potential energy can be tuned by the measurement and the feedback control. Consequently, it is possible to generate the quadrature squeezed state through this scheme.

We now turn to evaluate the second moments $\left\langle \hat{x}^2 \right\rangle $ and $\left\langle \hat{p}^2 \right\rangle $. The evolution equations of these two quantities take the following form:
\begin{align}
\frac{d \left\langle \hat{x}^2 \right\rangle }{dt} & = (k_2 - k_1)\left\langle \hat{x}^2 \right\rangle + \frac{k_1 + k_2}{2}
\notag
\\
& = -2 \kappa_f \left\langle \hat{x}^2 \right\rangle + \frac{\Gamma_p}{4},  \label{eq:x2}
\end{align}
\begin{align}
\frac{d \left\langle \hat{p}^2 \right\rangle }{dt} & = (k_2 - k_1)\left\langle \hat{p}^2 \right\rangle + \frac{k_1 + k_2 + 4k_3}{2}
\notag
\\
& = -2 \kappa_f \left\langle \hat{p}^2 \right\rangle + \frac{\Gamma_x}{4},  \label{eq:p2}
\end{align}
and the solutions of Eqs.~(\ref{eq:x2}) and (\ref{eq:p2}) are 
\begin{align}
\left\langle \hat{x}^2 \right\rangle & = C_1 e^{-2\kappa_f t} + \frac{ \Gamma_p }{8 \kappa_f}  
,
\label{eq:solutionx}
\\
\left\langle \hat{p}^2 \right\rangle & = C_2 e^{-2\kappa_f t} + \frac{ \Gamma_x }{8 \kappa_f},
\label{eq:solutionp}
\end{align}
where $C_1$ and $C_2$ are constants determined by the initial state of the system. Same with the averages $\left\langle \hat{x} \right\rangle$ and $\left\langle \hat{p} \right\rangle$ discussed before, the second moments, $\left\langle \hat{x}^2 \right\rangle$ and $\left\langle \hat{p}^2 \right\rangle$, converge to their own steady values exponentially, and the time scale for approaching the steady values is of the order of $\kappa_f^{-1}$.

The corresponding steady solutions of Eqs.~(\ref{eq:x2}) and (\ref{eq:p2}) are therefore
\begin{align}
\left\langle \hat{x}^2 \right\rangle & = \frac{ \Gamma_p }{8 \kappa_f}  
 = \frac{\gamma_p}{8\kappa_f} + \frac{\kappa_f}{2 \gamma_x},
\label{eq:uncerx}
\\
\left\langle \hat{p}^2 \right\rangle & = \frac{ \Gamma_x }{8 \kappa_f}  
= \frac{\gamma_x}{8\kappa_f} + \frac{\kappa_f}{2 \gamma_p}.
\label{eq:uncerp}
\end{align}
Here, we can see that for fixed $\kappa_f$, the uncertainty of $\hat{x}$ in the steady state monotonically increases with $\gamma_p$ but monotonically decreases with $\gamma_x$, and vice versa for the uncertainty of $\hat{p}$. This can be understood by the balance between the effects of the energy increment and reduction discussed in the previous section. In particular, the variances of two quadratures are equal when $\gamma_x = \gamma_p$, which has already been discussed in Ref.~\cite{A. Levy}. There, the authors have shown that the steady state of the system is an effective thermal state when the two measurement strengths are equal \cite{A. Levy}.

From Eqs.~(\ref{eq:uncerx}) and (\ref{eq:uncerp}), it is straightforward to obtain the uncertainty relation of $\hat{x}$ and $\hat{p}$ for the steady state:
\begin{align}
\left\langle \hat{x}^2 \right\rangle \left\langle \hat{p}^2 \right\rangle  & = \left(  \frac{\gamma_p}{8\kappa_f} + \frac{\kappa_f}{2 \gamma_x} \right) \left( \frac{\gamma_x}{8\kappa_f} + \frac{\kappa_f}{2 \gamma_p} \right)
\notag
\\
& = \frac{1}{8} + \frac{\gamma_x \gamma_p }{64 \kappa_f^2} + \frac{\kappa_f^2 }{4 \gamma_x \gamma_p}
\notag
\\
& \geqslant \frac{1}{8} + 2 \sqrt{\frac{\gamma_x \gamma_p }{64 \kappa_f^2} \frac{\kappa_f^2 }{4 \gamma_x \gamma_p}}
\notag
\\
& = \frac{1}{4}.   \label{eq:uncertainty}
\end{align}
The minimum-uncertainty relation given by the equality in the third line of Eq.~(\ref{eq:uncertainty}) holds under the following condition: 
\begin{align}
\gamma_x \gamma_p = 4\kappa_f^2.   \label{eq:relation}
\end{align}
Since the parameters $\gamma_x$, $\gamma_p$, and $\kappa_f$ are controllable by the observer, it is noted that as long as Eq.~(\ref{eq:relation}) is satisfied, there is still flexibility to control either $\left\langle \hat{x}^2 \right\rangle$ or $\left\langle \hat{p}^2 \right\rangle $ with keeping the ideal squeezing, $\left\langle \hat{x}^2 \right\rangle \left\langle \hat{p}^2 \right\rangle = 1/4$, by tuning the values of these three parameters. In addition, it has been proven that a state satisfying the minimal uncertainty relation of $\hat{x}$ and $\hat{p}$, $\left\langle \hat{x}^2 \right\rangle \left\langle \hat{p}^2 \right\rangle = 1/4$,  must be a pure state \cite{D. Stoler}. Therefore, our scheme allows us to produce any pure ideal quadrature-squeezed state with arbitrary squeezedness irrespective of the initial state.

In contrast, the feedback control based on the single observable measurement cannot generate an arbitrary quadrature-squeezed state. Without loss of generality, we assume that only $x$ quadrature is measured, and the feedback control Hamiltonian $\hat{H}'_f$  corresponding to Eq.~(\ref{eq:fb}) but for the single observable case is given by
\begin{align}
\hat{H}'_f dt \equiv  -\kappa_f \bar{x}(t)dt \cdot \hat{p}.  \label{eq:fb'}
\end{align}
Following the same approximation and calculation presented in the previous section, the ensemble averaged master equation under the single-observable measurement and the feedback control reads
\begin{align}
\frac{d \hat{\rho}}{dt} = & -i \left[ \hat{H}_s, \hat{\rho} \right] + k'_1 \mathcal{D}[\hat{c}] \hat{\rho} + k'_2 \mathcal{D}[\hat{c}^{\dagger}] \hat{\rho}  + k'_3 \mathcal{D}[\hat{c} + \hat{c}^{\dagger}] \hat{\rho} 
\notag
\\
& + k'_4(\hat{c} \hat{c} \hat{\rho}  + \hat{\rho} \hat{c}^{\dagger} \hat{c}^{\dagger} - \hat{c}^{\dagger} \hat{c}^{\dagger} \hat{\rho}  - \hat{\rho} \hat{c} \hat{c} ),  
\end{align}
with
\begin{align}
k'_1 & \equiv \frac{\kappa_f^2}{\gamma_x} + \frac{\kappa_f}{2}, 
\\
k'_2 & \equiv \frac{\kappa_f^2}{\gamma_x} - \frac{\kappa_f}{2},  
\\
k'_3 & \equiv \frac{\gamma_x }{8} - \frac{\kappa_f^2}{2\gamma_x}, 
\\
k'_4 & \equiv \frac{\kappa_f }{4}.
\end{align}
Again, ignoring the effect of the unitary evolution term, the resulting evolution equations of the first and the second moments of the two quadratures read
\begin{align}
\frac{d \left\langle \hat{x} \right\rangle }{dt} & = \frac{k'_2 - k'_1 - 4 k'_4}{2}\left\langle \hat{x} \right\rangle = -\kappa_f \left\langle \hat{x} \right\rangle,  \label{eq:x} 
\end{align}
\begin{align}
\frac{d \left\langle \hat{p} \right\rangle }{dt} & = \frac{k'_2 - k'_1 + 4 k'_4}{2} \left\langle \hat{p} \right\rangle = 0,
\label{eq:p} 
\end{align}  
and
\begin{align}
\frac{d \left\langle \hat{x}^2 \right\rangle }{dt} & = (k'_2 - k'_1 - 4 k'_4)\left\langle \hat{x}^2 \right\rangle + \frac{k'_1 + k'_2}{2}
\notag
\\
& = - 2\kappa_f \left\langle \hat{x}^2 \right\rangle + \frac{\kappa_f^2}{\gamma_x}, 
\label{eq:x^2}
\end{align}
\begin{align}
\frac{d \left\langle \hat{p}^2 \right\rangle }{dt} & = (k'_2 - k'_1 + 4 k'_4)\left\langle \hat{p}^2 \right\rangle + \frac{k'_1 + k'_2 + 4k'_3}{2}
\notag
\\
& =  \frac{\gamma_x}{4}, \label{eq:p^2} 
\end{align}
respectively. Equation (\ref{eq:x}) shows that the average of $x$ quadrature also reaches zero after a sufficiently long time for the single-observable measurement case, and Eq.~(\ref{eq:p}) shows that the average of $p$ quadrature is unchanged during the evolution. Consequently, the convergence and divergence of the uncertainties of two quadratures are exactly determined by the second moment of themselves, respectively. From Eqs.~(\ref{eq:x^2}) and (\ref{eq:p^2}), we can clearly see that the expectation value of the potential energy converges to a finite value while that of the kinetic energy is diverging after long time evolution. This means that by tuning the strengths of the measurement and the feedback control, the value of the uncertainty of the $x$ quadrature can be manipulated to be as small as possible, while the uncertainty of the $p$ quadrature is out of control. Thus, the arbitrary quadrature-squeezed state cannot be produced through this scheme based on the single observable measurement.

In addition, there are two other types of schemes proposed to prepare the quadrature-squeezed light using feedback control based on  the measurement of a single quadrature observable: one is mediated by the homodyne measurement, and the other is based on the quantum nondemolition measurement \cite{H. M. Wiseman}. For the former proposal, the source light is initially pumped to be squeezed, while the feedback control is performed to reduce the fluctuation of the unsqueezed quadrature. The quality of the squeezedness, of course, depends on the pump of the laser, and the minimum of the variance is half of that value of the coherent state for the perfectly regularly pumped laser. In addition, for the perfectly regular laser, the minimum of the uncertainty relation is around $20 \%$ greater than the lower bound of the Heisenberg uncertainty principle \cite{H. M. Wiseman}. In other words, neither the arbitrary nor the ideal quadrature squeezed state can be realized by this method. For the latter scheme, despite the fact that the variance of one quadrature can be squeezed to be arbitrarily small, that of the other unsqueezed quadrature is highly increased by the measurement \cite{H. M. Wiseman}. Consequently, the ideal squeezed state in general cannot be produced.

\section{numerical demonstrations \label{sec:results}}

In this section, we verify our results obtained in the previous section by numerical simulations. Then, we shall discuss experimental feasibility of our scheme.
\begin{figure}[tb!]
		
	\centering \includegraphics[scale=0.55]{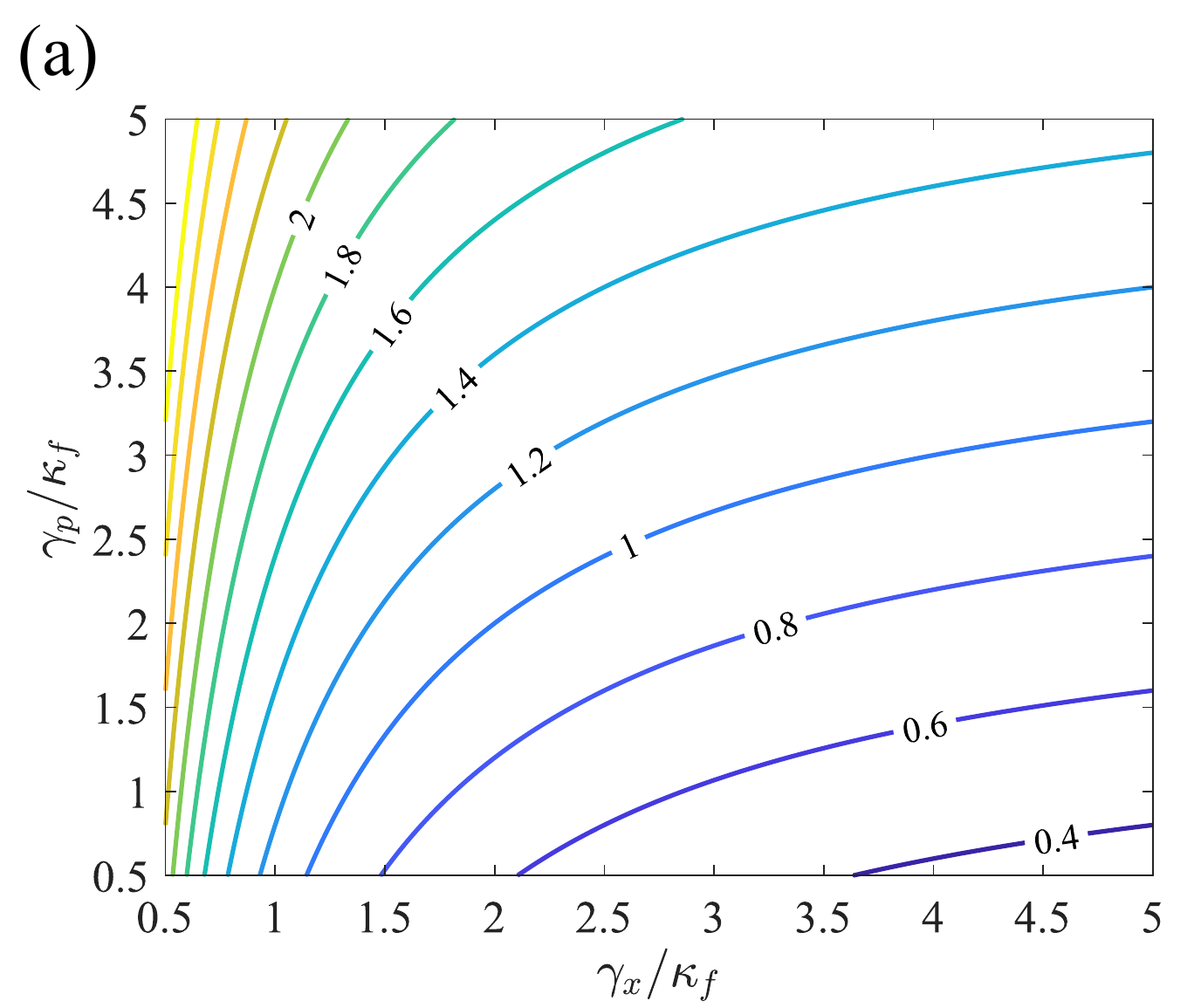}
	
	\centering \includegraphics[scale=0.53]{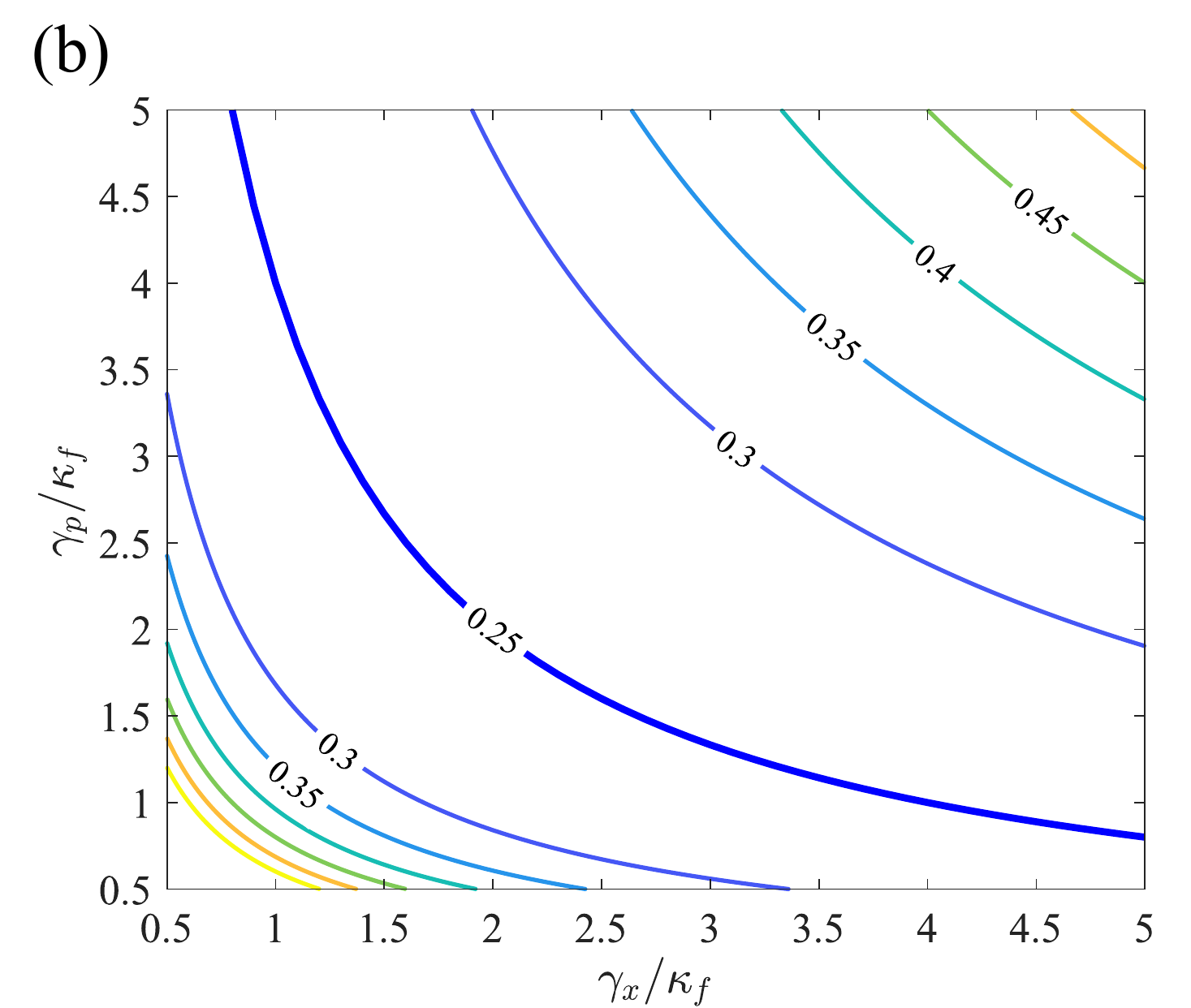}
	\caption{(a) The contour of $r_x$ with respect to $\gamma_x / \kappa_f$ and $\gamma_p / \kappa_f$. In the region where $r_x < 1$, the $x$ quadrature is squeezed. (b) The contour of $\left\langle \hat{x}^2 \right\rangle \left\langle \hat{p}^2 \right\rangle $ with respect to $\gamma_x / \kappa_f$ and $\gamma_p / \kappa_f$. The ideal squeezing, $\left\langle \hat{x}^2 \right\rangle \left\langle \hat{p}^2 \right\rangle = 1/4$, is realized when $\gamma_x \gamma_p = 4 \kappa_f^2$, which is shown by the thick line.}	
	\label{nounitary}
\end{figure}

\subsection{The properties of the steady state}

Let us first focus on the squeezedness of the steady state. It is convenient to introduce the following two parameters $r_x$ and $r_p$ to characterize the squeezing:
\begin{align}
r_x & \equiv \frac{\left\langle \hat{x}^2 \right\rangle}{1/2} = 2\left\langle \hat{x}^2 \right\rangle,
\\
r_p & \equiv \frac{\left\langle \hat{p}^2 \right\rangle}{1/2} = 2\left\langle \hat{p}^2 \right\rangle,
\end{align}
where $\left\langle \hat{x}^2 \right\rangle$ and $\left\langle \hat{p}^2 \right\rangle$ are normalized by their value, $1/2$, for the coherent state. The values of both $r_x$ and $r_p$ range from zero to infinity, and the value of either $r_x$ or $r_p$ smaller than unity represents that the state is quadrature squeezed. To better illustrate it, we take $r_x$ as an example and plot the contour of $r_x$ with respect to parameters $\gamma_x / \kappa_f$ and $\gamma_p / \kappa_f$ in Fig.~\ref{nounitary} (a). From this figure, one can observe that $r_x$ is monotonically increasing with parameter $\gamma_p / \kappa_f$ and monotonically decreasing with parameter $\gamma_x / \kappa_f$. For a given value of $\gamma_p / \kappa_f$, $r_x$ decreases to zero with increasing $\gamma_x / \kappa_f$, which means that we can generate squeezed states with arbitrarily high squeezing of the $x$ quadrature in principle. A similar discussion applies to the squeezing parameter $r_p$ for the $p$ quadrature as well, which concludes that we can generate squeezed states with arbitrarily high squeezing of the $p$ quadrature by taking sufficiently large $\gamma_p / \kappa_f$.

\begin{figure}[t!]
	\centering  \includegraphics[scale=0.56]{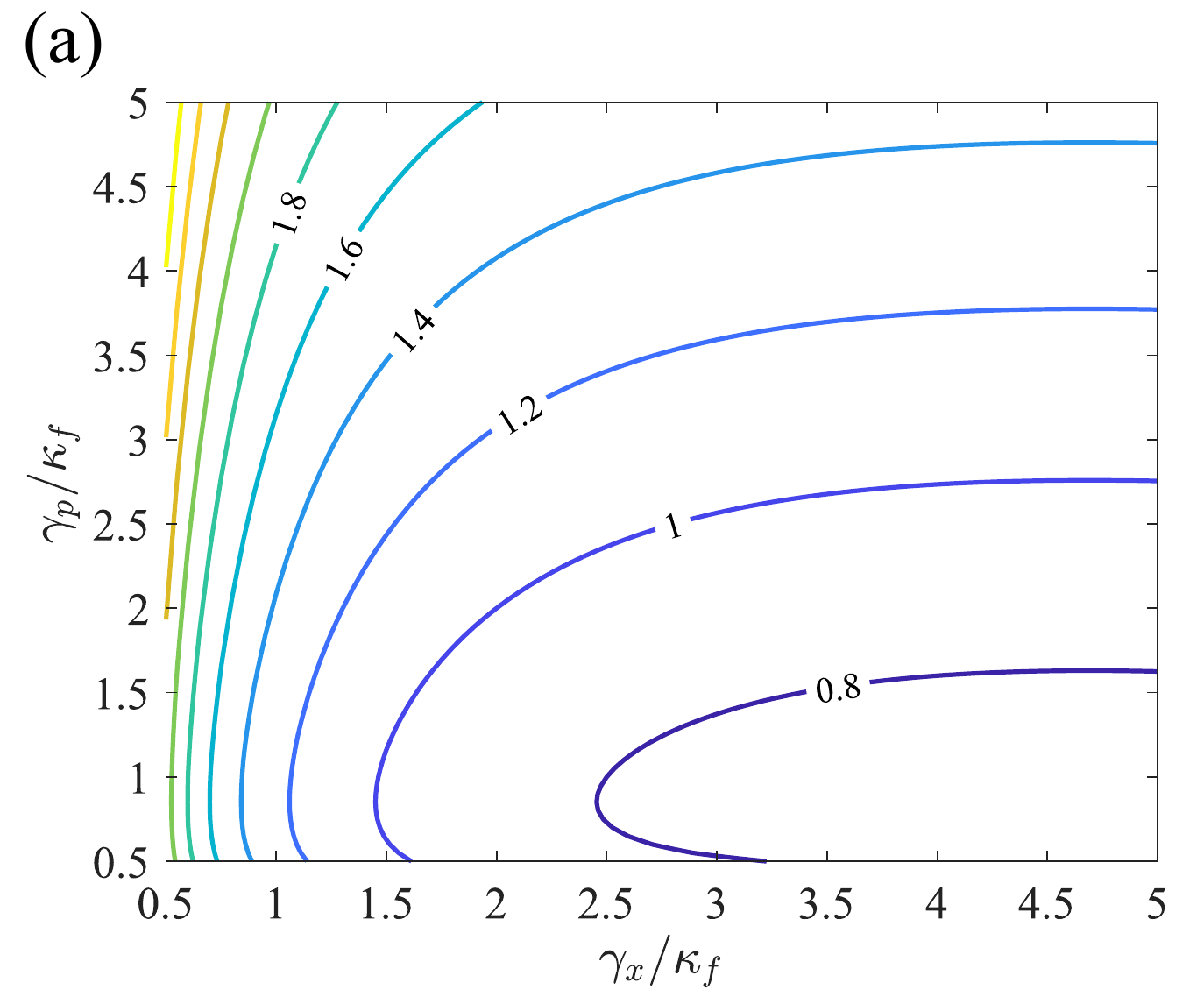}
	
	\centering \includegraphics[scale=0.56]{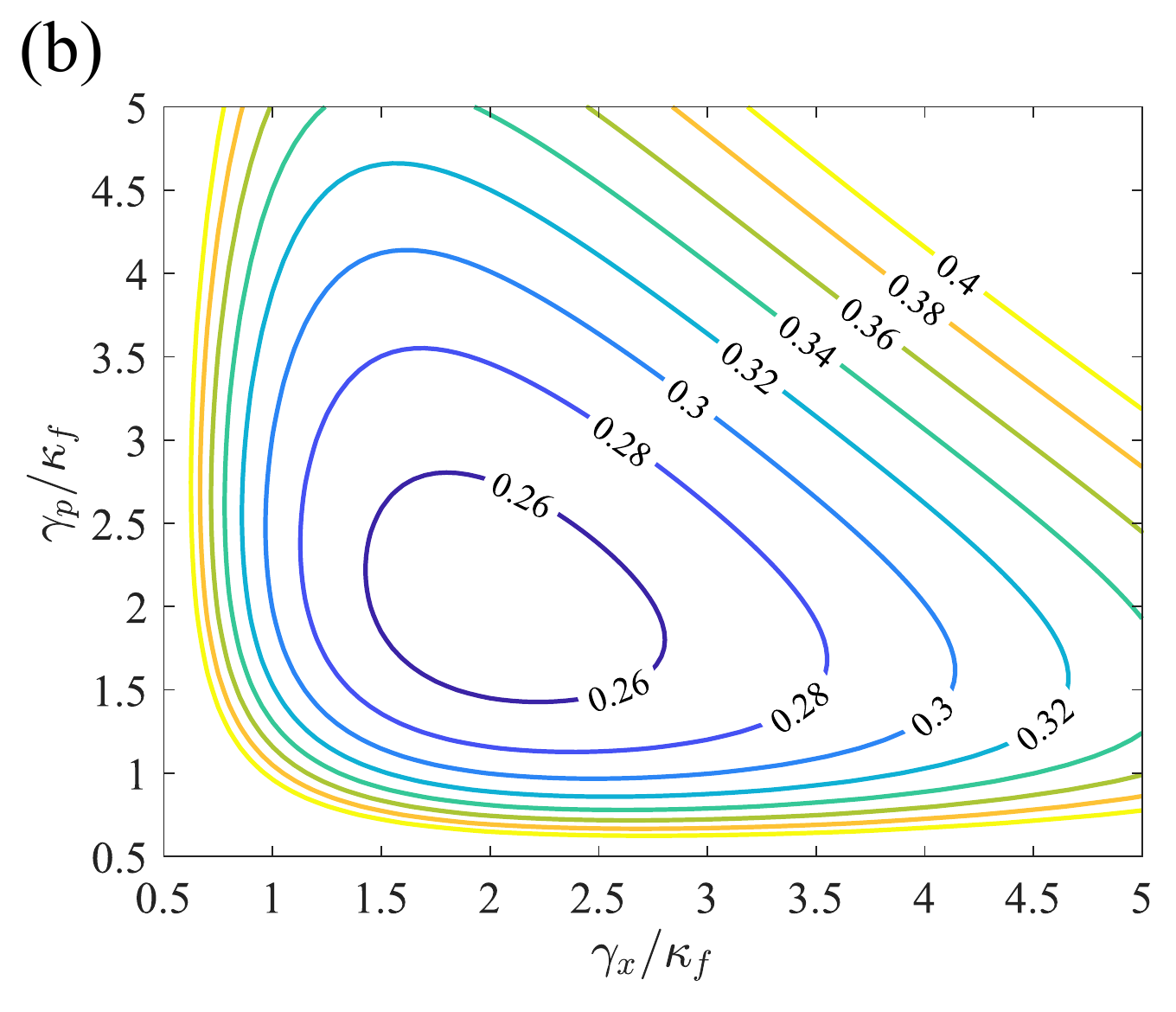}
	\caption{Same as Fig.~\ref{nounitary} but with the unitary evolution term, and setting $\kappa_f = 1.5$. (a) The contour of $r_x$ with respect to $\gamma_x$ and $\gamma_p$. (b) The contour of $\left\langle \hat{x}^2 \right\rangle \left\langle \hat{p}^2 \right\rangle $ with respect to $\gamma_x$ and $\gamma_p $. $\left\langle \hat{x}^2 \right\rangle \left\langle \hat{p}^2 \right\rangle = 1/4$ when $\gamma_x = \gamma_p = 2$.}	
	\label{unitary}
\end{figure}
Next, we consider the behavior of the quantity $\left\langle \hat{x}^2 \right\rangle \left\langle \hat{p}^2 \right\rangle $ in terms of the parameters $\gamma_x / \kappa_f$ and $\gamma_p / \kappa_f$ as shown in Fig.~\ref{nounitary}(b). As has been discussed in the previous section, $\left\langle \hat{x}^2 \right\rangle \left\langle \hat{p}^2 \right\rangle $ takes the minimum value $1/4$ along the line of $\gamma_p / \kappa_f = 4 (\gamma_x / \kappa_f)^{-1}$ or $\gamma_x \gamma_p = 4 \kappa_f^2$, which corresponds to the pure ideal quadrature-squeezed state. Together with Fig.~\ref{nounitary}(a), it is clearly seen that we can generate a pure ideal squeezed state with arbitrary high squeezing of the $x$ quadrature by taking sufficiently large $\gamma_x /\kappa_f$ and choosing $\gamma_p$ as $\gamma_p = 4 \kappa_f^2 /\gamma_x$. A similar argument holds for the $p$ quadrature as well. As a conclusion, any pure ideal quadrature-squeezed state can be generated by tuning $\gamma_x /\kappa_f$ and $\gamma_p /\kappa_f$ provided $\gamma_x \gamma_p = 4 \kappa_f^2$.

Finally, we complete our demonstration of the steady state by numerically showing the effect of the unitary evolution term. The parameter $\kappa_f$ is set to be $\kappa_f = 1.5$ as an example, and the remaining two parameters $\gamma_x$ and $\gamma_p$ are chosen as independent variables of order unity. In this case, $k_1$, $k_2$, and $k_3$ are comparable to unity, and the unitary evolution term should be taken into consideration. The numerical results of $r_x$ and $\left\langle \hat{x}^2 \right\rangle \left\langle \hat{p}^2 \right\rangle $ for the steady state as functions of $\gamma_x / \kappa_f$ and  $\gamma_p / \kappa_f$ are shown in Figs.~\ref{unitary}(a) and \ref{unitary}(b), respectively. Comparing Figs.~\ref{nounitary}(a) and \ref{unitary}(a), we find that the region to obtain the $x$ quadrature-squeezed state ($r_x < 1$) is narrowed down in the latter case with the unitary evolution term. Besides, for given values of $\gamma_x /\kappa_f$ and $\gamma_p /\kappa_f$ in this region, the squeezedness parameter $r_x$ becomes larger when we take account of the unitary evolution term. This means that the squeezedness is degraded. In addition, the ideal quadrature-squeezed state can no longer be generated: As can be seen from Fig.~\ref{unitary}(b), $\left\langle \hat{x}^2 \right\rangle \left\langle \hat{p}^2 \right\rangle$ takes its minimum value $1/4$ only when $\gamma_x / \kappa_f = \gamma_p / \kappa_f = 2$. However, from Fig.~\ref{unitary}(a), it can be observed that $r_x = 1$ at this point, namely the resulting state is a coherent state without squeezing. In summary, the effect of the unitary evolution term is to degrade the quality of the squeezedness.

\subsection{The behaviors during the evolution}

\begin{figure}[tb!]
		
	\centering \includegraphics[scale=0.55]{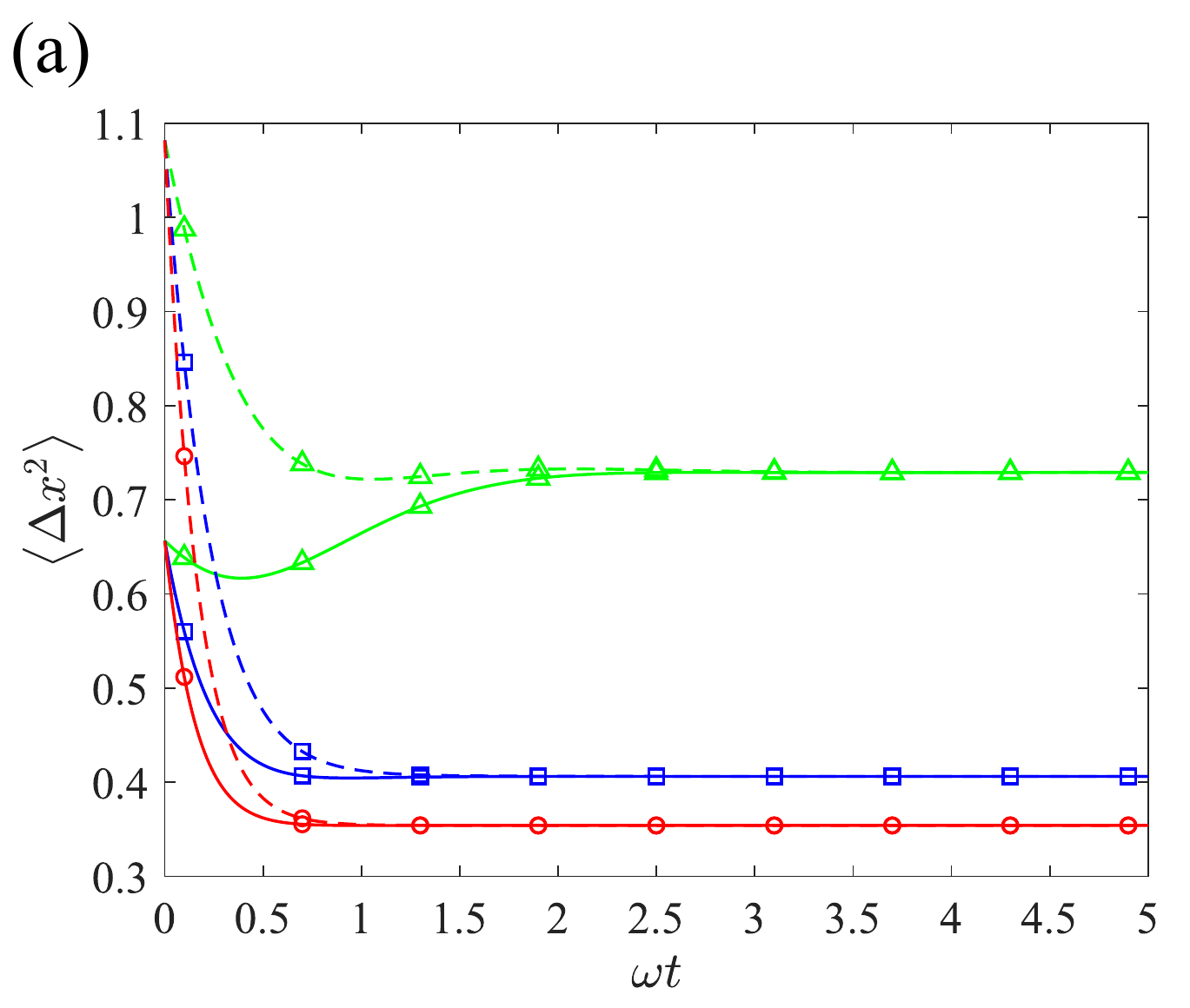}
		
	\centering \includegraphics[scale=0.55]{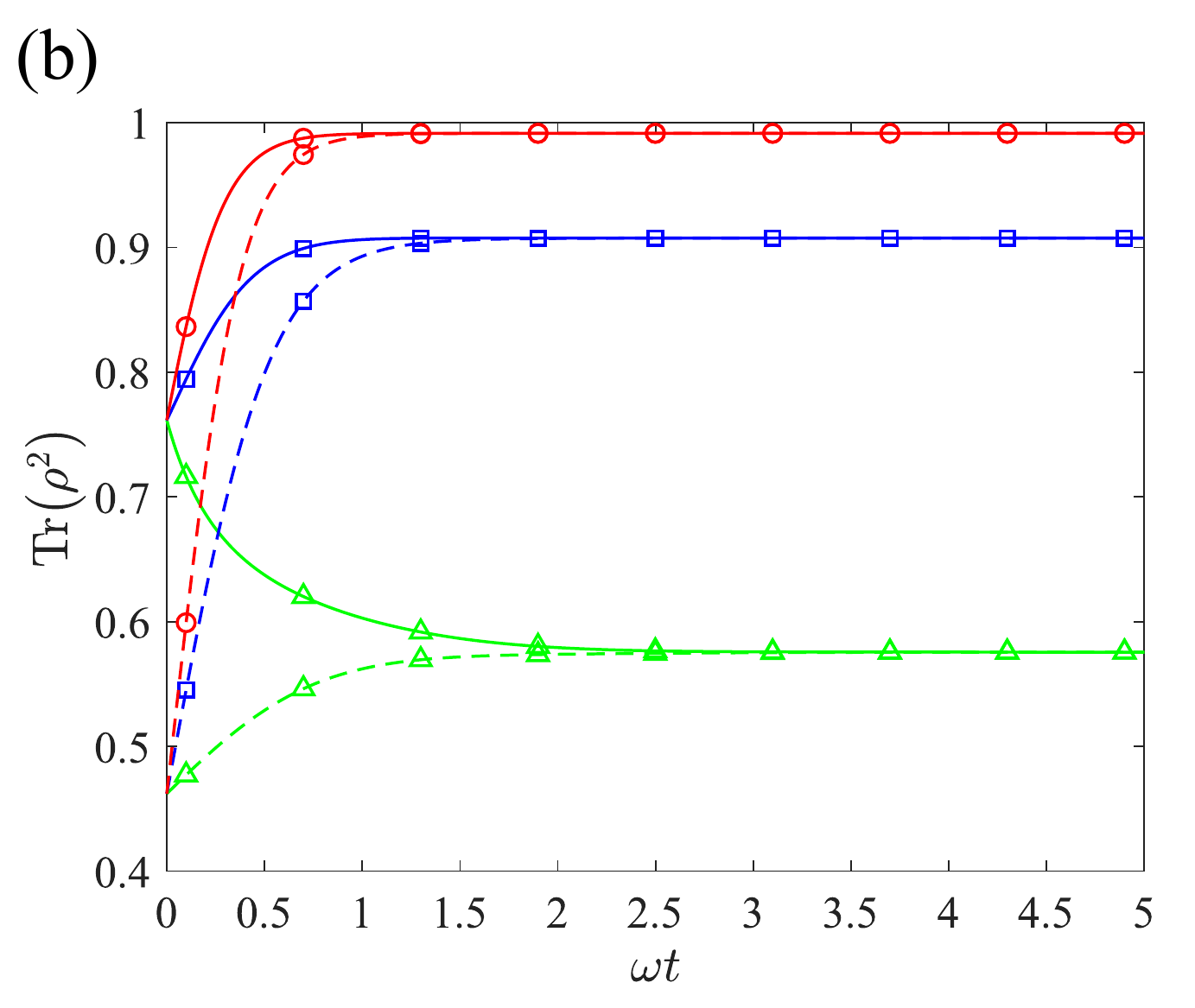}
	\caption{ Relaxation behavior of the variance $\left\langle \Delta x^2 \right\rangle $ and the purity $\Tr\left( \rho ^2 \right) $ to their steady value. (a) Time evolution of $\left\langle \Delta x^2 \right\rangle $.  (b) Time evolution of the purity $\Tr\left( \rho ^2 \right) $. The system is assumed to start from the thermal state with the inverse temperature $\beta = 1$ (dashed lines) and $2$ (solid lines). The green lines with triangles, blue lines with squares, and red lines with circles represent $\kappa_f = 1$, $2$, and $3$, respectively. The other parameters are $\omega = 1$, $\gamma_x = 9$, and $\gamma_p = 4$.} 	
	\label{evolution}
\end{figure}
The time scale to reach the steady state is of great concern in the practical experiments. Without loss of generality, here we take $x$ quadrature as an example, and the discussion of the $p$ quadrature can be made by following the same procedure. In Fig.~\ref{evolution}, we plot the time evolution of the variance of the $x$ quadrature and the purity for different initial states and the feedback control strengths $\kappa_f$. In this figure, the measurement strengths $\gamma_x$ and $\gamma_p$ are fixed while the feedback control strength $\kappa_f$ is varied since $\kappa_f$, as mentioned previously, plays an essential role in the time scale for reaching the steady state. For ease of comparison, the initial states are distinguished by different line types, and the feedback control strengths are distinguished by different colors and symbols. The initial state of the system is set to be a canonical state with the inverse temperature $\beta = 1$ (dashed lines) and $2$ (solid lines), and the measurement strength is set to be $\kappa_f = 1$ (green lines with triangles), $2$ (blue lines with squares), and $3$ (red lines with circles). From  Fig.~\ref{evolution}, we find that the time for reaching the steady state monotonically decreases with respect to $\kappa_f$ for a given initial state. When $\kappa_f$ is larger compared with $\omega$, this relaxation time, which is independent of the initial state, can be approximated by $\sim \kappa_f^{-1}$. Remarkably, when Eq.~(\ref{eq:relation}) is satisfied, and $\kappa_f$ is sufficiently greater than but still in the same order with $\omega$, we can almost approach to the results obtained in the previous section after sufficiently long time evolution. For instance, $\kappa_f = 3$, $\gamma_x = 9$, and $\gamma_p = 4$ (red lines with circles in Fig.~\ref{evolution}), the asymptotic value of the variance of the $x$ quadrature is $0.35$, which is close to the value ($\sim 0.33$) given by Eq.~(\ref{eq:uncerx}), and the purity of the density matrix eventually converges to $0.99$.

Finally, we discuss the experimental feasibility of our scheme taking a system of levitated particles in vacuum as an example. Levitated particles in vacuum is a promising platform for quantum sciences by virtue of its long coherence time and high precision of the control. Recently, Walker {\it et al.} proposed an experimentally feasible scheme to manipulate the motion of a levitated particle confined in a static magnetic trap by the measurement and the feedback control \cite{L. S. Walker}. Compared with an optical trap widely used in the current experiments, the magnetic trap gets rid of the problem of intrinsic laser recoil heating due to its low trap frequency ($\omega \sim 100 \ \rm Hz$) \cite{L. S. Walker, A. Rahman, Brien}. Thus, the reheating on the system mainly comes from the interaction with the environment. By putting the particle in the high vacuum ($10^{-10}$ mbar) or cooling the trap chamber cryogenically, this heating rate $\Gamma_{\rm th}$ can be dramatically reduced to the order of $1 \ \rm Hz$ \cite{L. S. Walker, V. Jain}. The feedback control strength $\kappa_f$ determines the relaxation rate of the system, whose typical value is around $10^3 \ \rm Hz$ in the experiments \cite{L. S. Walker, Ballestero}. Therefore, the system has already reached the target squeezed state before being reheated by the environment.

\section{Conclusion and future prospects  \label{sec:conclusion}}
 	
As an important application of our previous work \cite{C. Jiang}, we have proposed a theoretical scheme to produce the quadrature squeezed state of a harmonic oscillator system by the feedback control based on the simultaneous and continuous measurement of the noncommutative quadrature observables. Focusing on the asymptotic steady state, we have found that any pure ideal squeezed states can be generated by properly tuning the strengths of the measurements and the feedback control, which cannot be realized by the feedback control based on the single-observable measurement. Finally, we have demonstrated our conclusions by the numerical simulations.

Levitated microscopic systems have shown broad applications in various fields of physics \cite{J. Millen2, C. G.-Ballestero}. One of the most promising and inspiring applications is the ultrasensitive force detection, such as the precision measurement of the weak force (e.g., gravity and dispersion force) at short distances and that of weak force field. In recent years, great efforts have been devoted to realize such precision detection using the levitated particles as a sensor \cite{A. Kawasaki, C. P. Blakemore, D. Hempston, Z. Liu}. In the experiment, the motion of the center of mass of particles, such as spatial displacement or rotation, is measured as a response of the external force. Since the system is well isolated from the environment and the state of the particles is sensitive to the force, the measurement can be realized in a highly accurate level. In addition, the sensitivity of the detector could be further increased with an assist of the quadrature squeezedness of the oscillators \cite{D. Hempston, Z. Liu, E. E. Wollman, M. Rashid}, due to the enhancement of the resolution of the position of the particles. Another exciting application can be found in quantum thermodynamics. There, the novel phenomena introduced by the squeezed thermal reservoirs have been attracting increasing attention \cite{J. Rossnagel, G. Manzano, W. Niedenzu, J. Klaers1}. Recently, it is theoretically verified that the Landauer energy bound, i.e., the minimal energy for erasing $1$ bit information, can be exponentially reduced once the reservoir is in a quadrature squeezed state \cite{J. Klaers}. In the near future, squeezed levitated particles may offer an excellent platform for further experimental study of this issue. We hope our scheme will be used to prepare quadrature squeezed states in experiments of the above fields in the future.

\begin{acknowledgments}
We acknowledge Professor B. Prasanna Venkatesh for very helpful discussions and comments. This work was supported by NSF of China (Grants No.~11975199 and No.~11674283), the Zhejiang Provincial Natural Science Foundation Key Project (Grant No.~LZ 19A050001), the Fundamental Research Funds for the Central Universities (2017QNA3005, 2018QNA3004), and the Zhejiang University 100 Plan.
\end{acknowledgments}


\begin{thebibliography}{99}
\bibitem{note:noise} The It\^{o} rule implies that $d\xi_j$ can be treated as a random variable  of order $d t^{1/2}$.
\bibitem{note:term} The term $-i [\hat{H}_f dt, \hat{\rho}]$ is absent in the final master equation (\ref{eq:me}), since all the parts of this term either cancel with a part of the term $-i [\hat{H}_f dt, d \hat{\rho}]$ or vanish after taking the ensemble average.
\end{thebibliography}
\end{document}